\documentclass[pre,10pt,aps,twocolumn,amsmath,amssymb,superscriptaddress]{revtex4-1}
\pdfoutput=1
\usepackage{hyperref}
\usepackage{graphicx}
\usepackage{color}
\usepackage{multirow}
\usepackage{amsmath}
\usepackage{array}
\usepackage{booktabs}
\usepackage{multirow}
\usepackage{cleveref}

\newcommand{\be}{\begin{equation}}
\newcommand{\ee}{\end{equation}}
\newcommand{\bea}{\begin{eqnarray}}
\newcommand{\eea}{\end{eqnarray}}
\def\c#1{~\cite{#1}}
\def\f#1{Fig.~\ref{#1}}

\def\beq{\begin{equation}}
\def\eeq{\end{equation}}

\begin{document}

\title{Finite-dimensional vestige of spinodal criticality above the dynamical glass transition}

\author{Ludovic Berthier}

\affiliation{Laboratoire Charles Coulomb (L2C), Universit\'e de Montpellier, CNRS, 34095 Montpellier, France}

\affiliation{Department of Chemistry, University of Cambridge, Lensfield Road, Cambridge CB2 1EW, United Kingdom}

\author{Patrick Charbonneau}

\affiliation{Department of Chemistry, Duke University, Durham, North Carolina 27708, USA}

\affiliation{Department of Physics, Duke University, Durham, North Carolina 27708, USA}

\author{Joyjit Kundu}
\affiliation{Department of Chemistry, Duke University, Durham, North Carolina 27708, USA}

\email{joyjitkundu032@gmail.com}

\date{\today}

\begin{abstract}
Finite-dimensional signatures of spinodal criticality are notoriously difficult to come by. 
The dynamical transition of glass-forming liquids, first described by mode-coupling theory, is a spinodal instability preempted by thermally activated processes that also limit how close the instability can be approached. 
We combine numerical tools to directly observe vestiges of the spinodal criticality in finite-dimensional glass formers. We use the swap Monte Carlo algorithm to efficiently thermalise configurations beyond the mode-coupling crossover, and analyze their dynamics using a scheme to screen out activated processes, in spatial dimensions ranging from $d=3$ to $d=9$. We observe a strong softening of the mean-field  square-root singularity in $d=3$ that is progressively restored as $d$ increases above $d=8$, in surprisingly good agreement with perturbation theory. 
\end{abstract}

\maketitle

\section{Introduction}Spinodals predicted by mean-field theories do not exist in finite-dimensional systems because thermal (or other) fluctuations destabilize the precursor metastable state before the critical point can ever be reached~\cite{Gunton1978,Gunton1983,Ray1990,Stillinger2015}. An Ising system prepared with the metastable magnetization, for instance, grows nuclei (instantons~\cite{Langer1967,Langer1969}) of the opposite magnetization well ahead of the formal instability region. As the single-flip relaxation dynamics slows down critically, that of nucleation accelerates\c{ray1991}; cluster moves only worsen the imbalance. Numerical studies have nevertheless observed convincing hints of the Ising spinodal pseudocriticality in high enough spatial dimension $d$. Nucleation kinetics being exponentially suppressed as $d$ increases, the pseudocritical power-law scaling can then be made sufficiently extended\c{ray1991}. 

More theoretically enticing than the Ising spinodal is that of models with disorder, which capture the essence of systems ranging from magnetic\c{sethna1993} to mesoporous\c{aubry2014,tarjus2004} materials and also appear in social sciences and economics\c{soc2013}. These models exhibit a rich set of activated processes, such as avalanches\c{sethna1993,sethna1999} and hopping\c{patrick_pnas2014,foot:1}, in addition to nucleation, which makes their criticality especially challenging to scrutinize. Hence, although the Ginzburg criterion for the corresponding cubic field theory without activation gives an upper critical dimension $d_\emph{u}=8$\c{biroli2007,franz2011,zamponi2012}, it is unclear how relevant the associated pseudocriticality might be in any given system. Below $d_\emph{u}$, perturbative expansions relying on dimensional reduction~\cite{franz2011,zhong2017} or direct expansion~\cite{Rizzo2016} further do not concur. Even if they did, one may expect non-perturbative fluctuations to also contribute~\cite{saroj2016}. Although these fluctuations should limit the relevance of perturbative approaches and question the very existence of an upper critical dimension, very little is known about their effects~\cite{lubchenko2003barrier,bhattacharyya2008facilitation}, emphasizing further the need for quantitative results in finite-dimensional systems.  

This theoretical haze has not held back the use of mode-coupling theory (MCT) to describe the dynamics of supercooled liquids\c{mct_book}, which has been the subject of countless numerical and experimental tests\c{mct_book,gotze1999recent}. Although unclear in its initial derivations, it is now understood that MCT~\cite{andreanov2009mode} as well as the mean-field $d\rightarrow\infty$ description of liquids~\citep{Charbonneau:2017} indeed correspond to the limit of stability of the glass phase upon heating (or decompressing). The dynamical glass transition should thus be described as a spinodal (thermodynamic) instability in the presence of quenched disorder~\cite{franz1997phase,saroj2016}. Unfortunately, the spinodal is found to be totally hidden by finite-dimensional effects in direct free energy calculations~\cite{berthier2013overlap,sho_pnas}. 
 
The associated critical scaling laws of the structural relaxation time $\tau_\alpha$ and of time correlation functions upon approaching the avoided MCT (dynamical) transition from the equilibrium liquid are instead much more frequently examined\c{mct_book,gotze1999recent,Charbonneau:2017}. These quantities are straightforwardly measured in both simulations and experiments, but they are also non-universal, i.e., model dependent. Theoretical predictions for the associated critical exponents are not only sensitive to the spatial dimension (even above $d_\mathrm{u}$) and to activated processes, but also to fine details of the liquid structure and pair interactions. These predictions are thus typically of limited quantitative validity in finite $d$, but this limitation did not prevent MCT from making valid predictive statements regarding the glassy dynamics of a variety of materials~\cite{mct_book,gotze1999recent}.

In addition to non-universal scaling laws, the spinodal criticality is associated to a few universal signatures. In particular,  a square root singularity of the Edwards-Anderson parameter directly follows\c{mct_book,zamponi_rmp}, which is dynamically accessible as the plateau height in time correlation functions or the typical cage size in particle displacements. Treating fluctuations beyond mean-field leads also to universal predictions regarding the behaviour of four-point susceptibilities\c{glotzer2002,andreanov2009mode,berthier2007spontaneous}. Yet, because of the computational difficulty of equilibrating liquids beyond the avoided dynamical transition, and of the lack of experimental methods to screen out activated processes, it remains difficult to assess signatures of the spinodal, even in simple glass-forming liquids. As a result, the validity of the square root singularity remains a debated issue~\cite{berthier2007revisiting}. 

In this work, we exploit a recent implementation of swap Monte Carlo (SWAP) for continuously polydisperse systems, which bypasses the sluggishness associated with approaching the avoided dynamical transition\c{swap-prl,ludo_prx,kundu2019}, in order to probe the spinodal criticality beyond the MCT crossover in equilibrium conditions. By carefully controlling for activated processes, our analysis manages to extract a sufficiently long scaling regime of the typical cage size to estimate effective critical exponents and the putative presence of the square-root singularity across several space dimensions from the experimentally relevant $d=3$, where we conclude that the singularity is considerably softened, up to $d=9$ where a nearly perfect square-root scaling can be convincingly observed. 

\section{Simulation details}We consider a continuously polydisperse system of $N$ hard spheres under periodic boundary conditions in a simulation box of constant volume $V$. A hypercubic box is used in $d=3,4,5,6$, and $8$, while in $d=7$ and in $d=9$ we use the Wigner-Seitz cell of the checkerboard lattice 
that decreases the number of simulated particles while preserving the same effective box size. The particle-size distribution is $P(\sigma)=K/\sigma^3$, with normalization constant $K$ for diameters $\sigma \in \{\sigma_{\rm max},\sigma_{\rm min}\}$ where $\sigma_{\rm max}$, $\sigma_{\rm min}$ are the maximum and minimum diameters for a given polydispersity. The average diameter ${\bar \sigma}$ sets the unit of length, the degree of polydispersity is defined by the standard deviation of the diameter distribution, and the packing fraction is $\varphi=\rho \bar{V}_d$ for a number density $\rho=N/V$ and average volume of a $d$-dimensional hypersphere $\bar{V}_d$. The degree of polydispersity is chosen to be the minimum needed for the SWAP efficiency to saturate, i.e., $23\%$ for $d=3$, $10\%$ for $d=4-8$, and $8\%$ for $d=9$\c{kundu2019}, and suitably optimized SWAP sampling is used to equilibrate initial configurations. This approach ensures fast structural relaxation without crystallization or fractionation. 
Structural equilibration is notably validated by the complete decay of the self-part of the particle-scale overlap function
\begin{equation}
Q(t)=\frac{1}{N}\displaystyle \sum_{i=1}^{N} \Theta(a-|\mathbf{r}_i(t)-\mathbf{r}_i(0)|),
\end{equation} 
where $\Theta$ is the Heaviside function and $a=0.3\bar{\sigma}$ is about the typical particle cage size\c{ludo_prx,sho_pnas,sho_nat,kundu2019}. The associated structural relaxation time, $\tau_\alpha$, is defined as $Q(\tau_\alpha)=1/e$. From these initial equilibrium configurations, multiple simulations are then run with a purely local Monte Carlo dynamics. This computational scheme achieves equilibration at densities $3-8\%$ above the estimated dynamical transition, depending on $d$\c{kundu2019,ludo_prx}. This strategy opens a comfortable regime to study glassy dynamics approaching the mode-coupling crossover from the arrested phase, unavailable to previous computational work. 

\begin{figure}
\includegraphics[width=\columnwidth]{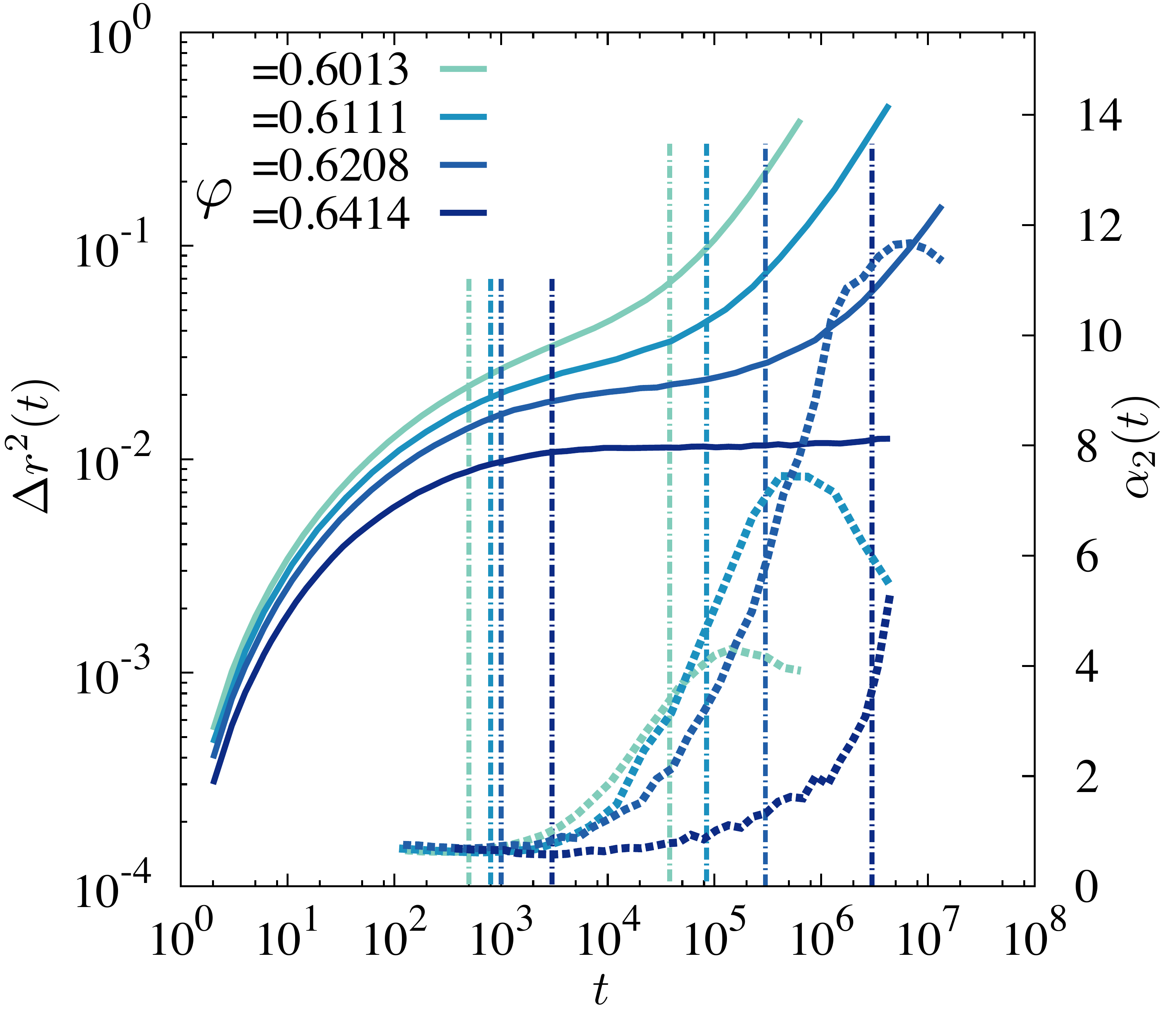}
\caption{Time evolution of the MSD $\langle \Delta r^2(t)\rangle$ and of the non-Gaussian parameter $\alpha_2(t)$ for different densities in $d=3$.  Vertical lines denote the start and end times of the window over which the cage size is measured for different densities.}
\label{fig01}
\end{figure}

\begin{table*}
\begin{tabular}{|p{1.00cm}|p{1.25cm}|p{1.25cm}|p{1.25cm}|p{1.25cm}|p{1.25cm}|p{1.45cm}|p{1.45cm}|}
 \hline
 $d$ & 3 & 4 & 5 & 6 &7 & 8 & 9\\
  \hline
 $\varphi_{\rm d}$& $0.600 (2)$ & $0.410(2)$ & $0.277 (1)$ & $0.1808(8)$& $0.1147 (6)$ & $0.0716 (3)$ & $0.0426 (2)$ \\
  \hline
 $\hat\Delta_d$& $0.015 (2)$ &  $0.021 (2)$& $0.020(2)$ & $0.018 (1)$ & $0.016(1)$ & $0.010 (1)$ & $0.013(1)$ \\
  \hline
 $\hat\Delta(\varphi_{\rm d})$& $0.0148 (5)$ &  $0.0208 (5)$& $0.0196(5)$ & $0.0175 (6)$ & $0.0161 (6)$ & $0.0101 (4)$ & $0.0129 (5)$ \\
 \hline
$1/\delta$& $0.74 (9)$ & $0.68(9)$ & $0.65 (8)$ & $0.61 (9)$ & $0.59(9)$ & $0.53 (9)$ & $0.55(10)$\\
 \hline
$A_d$& $0.11 (2)$ & $0.16 (2)$ & $0.15 (2)$ & $0.12(2)$ & $0.11(2)$ & $0.04 (1)$ & $0.09 (3)$\\
 \hline
\end{tabular}
\caption{Fit parameters $A_d$ and $\hat\Delta_d$ for Eq.~\eqref{cage_size_scaling} for different dimensions. The results for $\varphi_\mathrm{d}$ are obtained by standard MCT dynamical scaling~\cite{Charbonneau:2017}. The direct evaluation of the typical cage size at the estimated $\varphi_\mathrm{d}$ validates the value of the fitted quantity. Error bars on the fit parameters are determined from quality of the fit ($R^2$) to Eq.~\eqref{eq:cage}.  
 }
 \label{table01}
\end{table*}

\section{Typical cage size}We first consider the size of the typical cage, $\hat{\Delta}$, which in the MCT and the mean-field description of hard spheres is expected to scale as
\bea
{\hat \Delta (\varphi)}&=&{\hat \Delta_{\rm d}}-A_d(\varphi-\varphi_{\rm d})^{1/\delta}, 
\label{eq:cage}
\eea
with $1/\delta=1/2$ and $\hat \Delta_{\rm d}=\hat\Delta(\varphi_\mathrm{d})$ for densities beyond the dynamical transition, $\varphi>\varphi_{\rm d}$. In the high-dimensional limit, $\hat{\Delta}$ could be extracted from the long-time plateau of the mean square displacement (MSD) of an individual particle, $\hat{\Delta}=\Delta r_i^2(t)=\int r^2 G_{\rm s}(\mathbf{r},t) d\mathbf{r} $, where $G_\mathrm{s}(\mathbf{r},t)$ is the self-part of the van Hove function. In finite dimensions, two difficulties arise. First, caging is heterogeneous, and hence the full distribution of cages must be considered. Second, a sharp plateau in the MSD can only be identified much beyond the (avoided) dynamical transition, i.e. too far beyond the regime of interest. We consider the second effect first. Because activated processes interfere with the formation of the MSD plateau, the size of the transient cage must be extracted over a finite time window. In order to identify the mean-field-like dynamical caging regime, we rely on the non-Gaussian parameter, $\alpha_2(t)=\frac{d}{d+2}\frac{\langle r^4(t)\rangle}{\langle r^2(t)\rangle^2}-1$, as illustrated in \f{fig01}. More specifically, the upper bound of the time window is set at $20\%$ of peak non-Gaussianity, i.e., the maximum of $\alpha_2(t)$, and the lower bound is set to the short-time plateau of $\alpha_2(t)$. Because the lower bound is much smaller than the upper bound, results are insensitive to its precise value, while the upper bound only weakly affects the subsequent analysis as long as it is chosen consistently. Note that for packing fractions much above $\varphi_\mathrm{d}$ (in $d=3$, $\varphi \gtrsim 0.635$), the system does not relax within the simulation timescale (with standard dynamics), which results in a clear plateau in the MSD (\f{fig01}). The upper bound is then even tighter. Note also that generalizing the MSD to further suppress the contribution of activated processes as in Ref.~\c{patrick_pnas2014} markedly flattens the MSD, but does not quantitatively affect the subsequent analysis~(see Appendix~\ref{sec:gmsd}. 

\begin{figure}[b]
\includegraphics[width=\columnwidth]{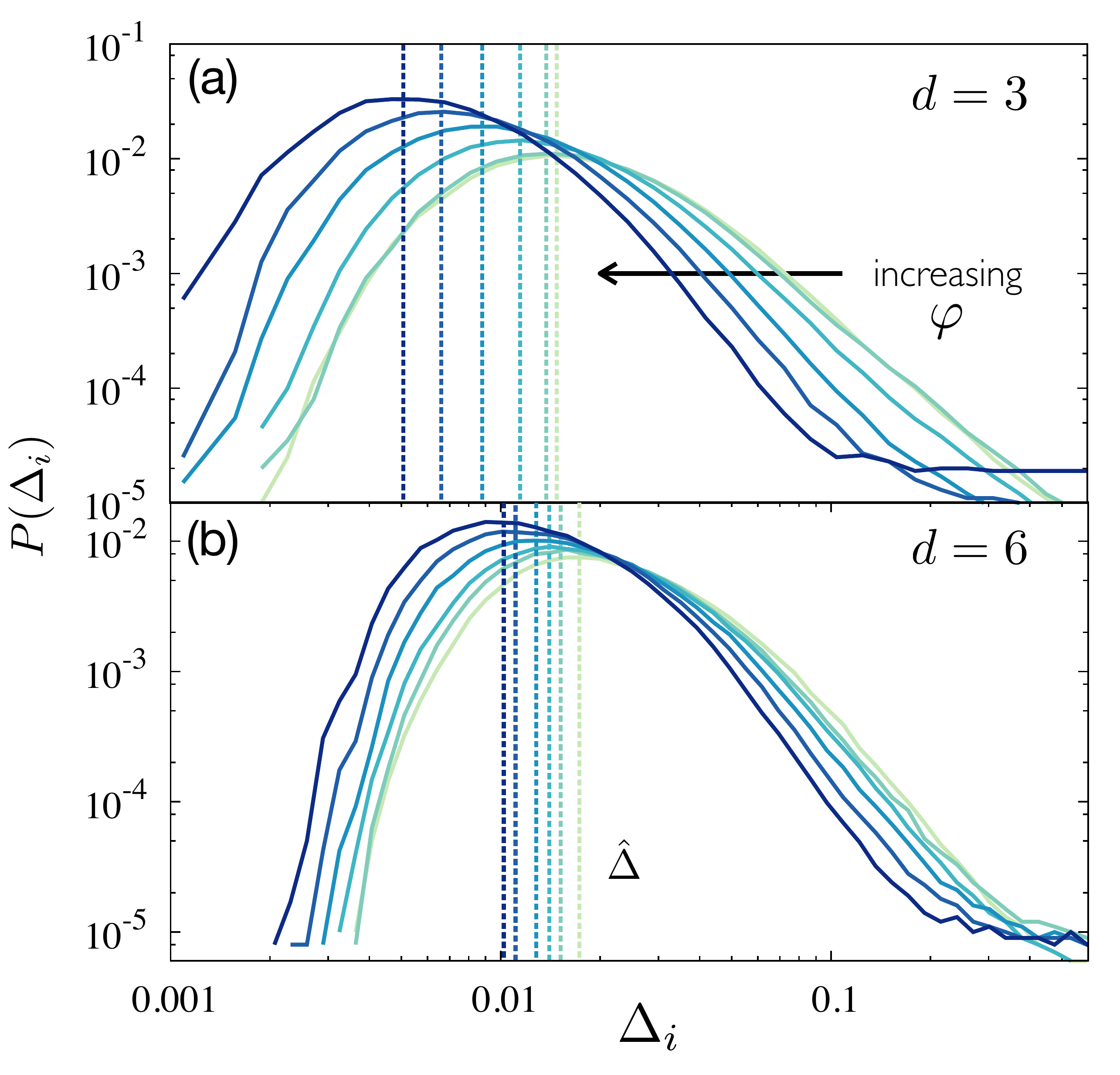}
\caption{Cage size distributions, $P(\Delta_i)$, in (a) $d=3$ for $\varphi=0.6005, 0.6032, 0.6111, 0.6208, 0.6309$, and $0.6414$ and (b) $d=6$ for  $\varphi=0.1810$, 0.1829,  0.1842, 0.1865, 0.1892, and 0.1916. Note that $\varphi_{\rm d}=0.600(2)$ and $0.1808(8)$ in $d=3$ and $6$ respectively. Fat tails at large displacements persist over the whole density regime accessible in simulations. The estimator $\hat \Delta=\mathrm{argmax}_{\Delta_i} P(\Delta_i)$ (vertical dashed lines) nonetheless monotonically shifts to smaller values as $\varphi$ increases.}
\label{cage_size_dist}
\end{figure}

\begin{figure}[b]
\includegraphics[width=\columnwidth]{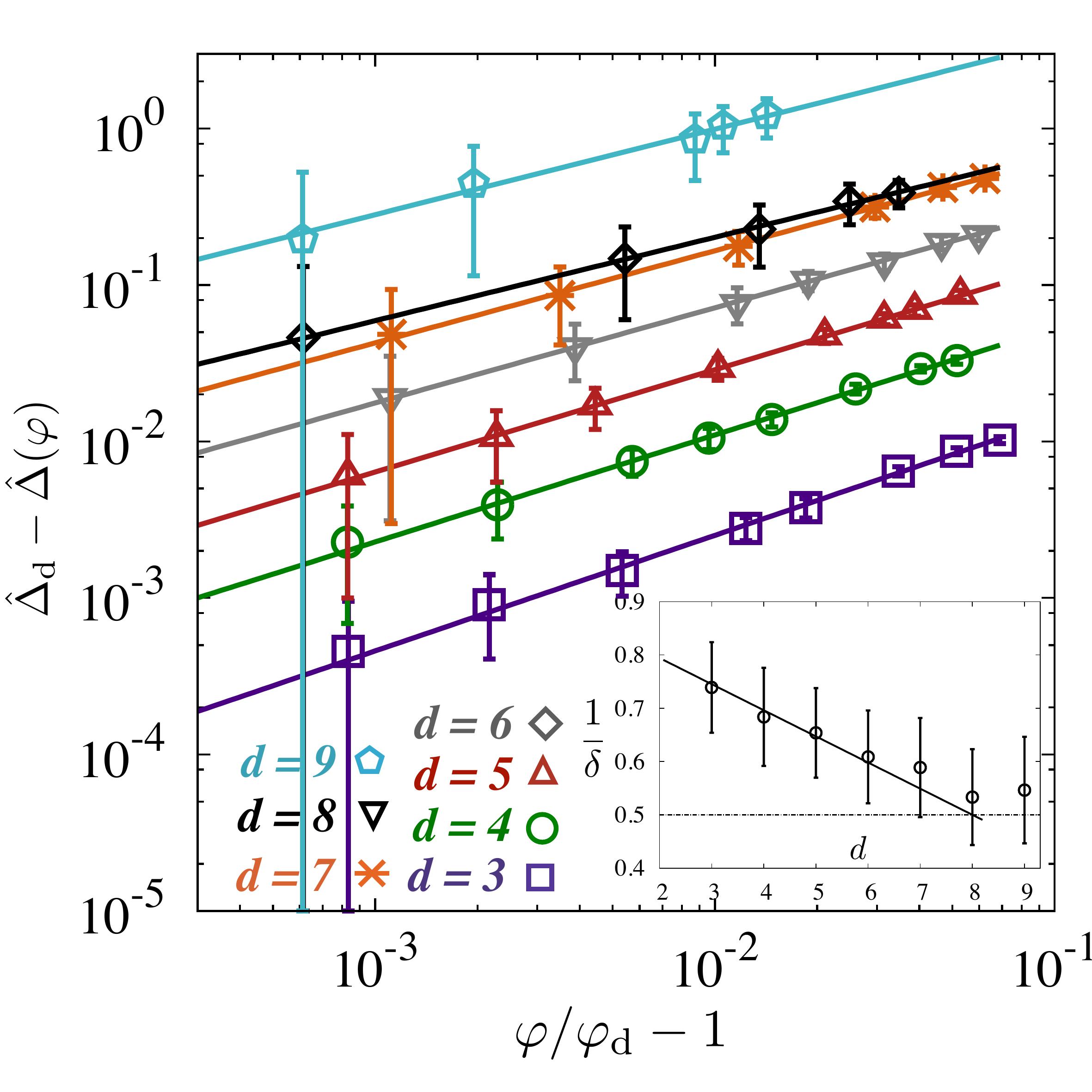}
\caption{Critical scaling of the typical cage size in $d=3\ldots 9$ as well as (inset) fitted power-law exponents, $1/\delta$. For visual clarity, data is vertically shifted by a factor of $3^{d-3}$. Error bars reflect the measurement uncertainty of $\Delta(\varphi)$ only with $\Delta_d$ and $\varphi_{\rm d}$ here chosen to optimize the quality ($R^2$) of the fit. Inset: the critical exponent $1/\delta$ versus $d$. A clear change in critical scaling is seen around the upper critical dimension, $d_\mathrm{u}=8$, predicted from perturbative approaches.  For $d>d_\mathrm{u}$, the results are consistent with the mean-field prediction ($1/\delta=1/2$, dashed line); for $d<d_\mathrm{u}$ the deviation from the mean-field result is approximately linear $1/\delta-1/2=B (d_\mathrm{u}-d)$ with $B=0.049(3)$.}
\label{cage_size_scaling}
\end{figure}

Once the timescale for identifying the cage dynamics is set, we can measure the cage size for each particle, $\Delta_i$, and average over samples and particles to define the cage size distribution $P(\Delta_i)$.  
The results in \f{cage_size_dist} show that cages generally tighten as density increases in all dimensions, but that fat tails at large displacements  persist for all densities. These tails deviate significantly from the log-normal forms reported in some mean-field models\c{patrick_pnas2014}. Although relatively little theoretical guidance is available as to what the proper functional form for $P(\Delta_i)$ should be~\cite{Chaudhuri2007}, our observations suggest that activated processes are not fully eliminated from the MSD analysis. To further sidestep this issue, we use as estimator of the typical cage size the mode of the distribution, $\hat{\Delta}=\mathrm{argmax}_{\Delta_i} P(\Delta_i)$, which is much less sensitive to the activated processes that appear in the fat tail of the distribution than the mean cage size, but converges to the same quantity as $d\rightarrow\infty$. 

In order to assess the critical scaling of the typical cage size, $\hat\Delta(\varphi)$ is fitted to Eq.~\eqref{eq:cage} using $\delta,$ $\hat\Delta_d$ and $A_d$ as parameters, while $\varphi_{\rm d}$ is obtained independently from the growth of the relaxation time $\tau_\alpha(\varphi)$\c{kundu2019} (Fig.~\ref{cage_size_scaling}, Table{\ref{table01}). Because of the uncertainty on $\varphi_\mathrm{d}$, the cage size at $\varphi_{\rm d}$, is not directly measured; but the fitted value is consistent with the direct estimate, which validates our approach. The values of the fit parameters and direct measurements are listed in Table~\ref{table01}.  
Note that both the fit error on $\Delta_{\rm d}$ at a given $\varphi_{\rm d}$ and the propagated uncertainty from $\varphi_{\rm d}$ are then included. 
 In contrast to earlier (cruder) estimates\c{patrick_pnas2012}, we find that for a given polydispersity,  $\hat\Delta_d$ decreases monotonically with increasing dimension\c{zamponi_rmp}. More significantly, we also find that $1/\delta$ decreases monotonically, and nearly linearly, as dimension increases, from $0.74(9)$ in $d=3$ down to values numerically indistinguishable from the mean-field prediction, $1/\delta = 1/2$, around $d\geq8$. 
Although the precise numerical estimates of $\delta$ are fairly robust to the details of the above analysis, they may still be fragile to the overall scheme. The detection of the upper critical dimension in the vicinity of $d=8$, and the systematic softening of the square-root singularity below $d_\mathrm{u}=8$ are nonetheless numerically robust (see Appendix~\ref{sec:gmsd} and \ref{sec:finitesize}). Interestingly, the latter is in sharp contrast with the prediction from dimensional reduction~\cite{franz2011,zhong2017} .

\section{Conclusion}The finite-dimensional vestige of the spinodal criticality associated with the dynamical transition of glass forming liquids has here been characterized by numerical simulations using the SWAP algorithm and a careful screening of activated processes across a broad range of spatial dimensions. Our simulations reveal that the square-root singularity, often used to describe experimental measurements in molecular glass formers, is dramatically softened in $d=3$ for hard spheres, a canonical model for testing MCT predictions. The measured effective exponent, $1/\delta \approx 0.75$, can still be considered as indirect evidence for an underlying avoided singularity, because $1/\delta=1$ would be trivially expected for a featureless evolution of the cage size. Further, the slow variation of $1/\delta$ towards $1/2$ with spatial dimension $d$ suggests that strong deviations from mean-field criticality exist even in large spatial dimensions. It takes simulations in dimensions $d\geq8$ to observe direct signatures of the square-root scaling that underlies the mean-field dynamical glass transition. The agreement between our results with perturbative approaches is nevertheless surprising, given the expected role of nonperturbative physics in the avoidance of the dynamical glass transition~\cite{saroj2016}, which can in principle persist even above $d=8$. A possible explanation might be the relative insensitivity of our specific estimators to these effects. 

Because this spinodal critical point is part of a broad universality class~\cite{saroj2016}, we expect our results to apply to a variety of other systems, for which the interplay between activation and criticality might be harder to control. Most crucially, these results further motivate the use of the dynamical criticality and of the mean-field description in describing the behavior of finite-dimensional liquid glass formers. It was recently shown that deviations from the dynamic glass transition can be studied considering the degree of localization of unstable modes in the potential energy landscape\c{daniele2018}. Localized excitations are indeed expected to disappear as $d$ increases, and our study thus also motivates verifying this prediction in larger dimensions.

\begin{acknowledgments}
We thank G.~Biroli, G.~Carra, S.~Franz, G.~Tarjus, and F.~Zamponi for stimulating discussions, as well as Tom Milledge for technical support.  We acknowledge support from the Simons Foundation grant (\#454933, LB, \# 454937, PC). Simulations were performed at Duke Compute Cluster (DCC), and at the Extreme Science and Engineering Discovery Environment (XSEDE), supported by National Science Foundation grant number ACI-1548562. 
\end{acknowledgments}

\appendix
\section{Generalized mean square displacement}
\label{sec:gmsd}
\begin{figure}
\includegraphics[width=\linewidth]{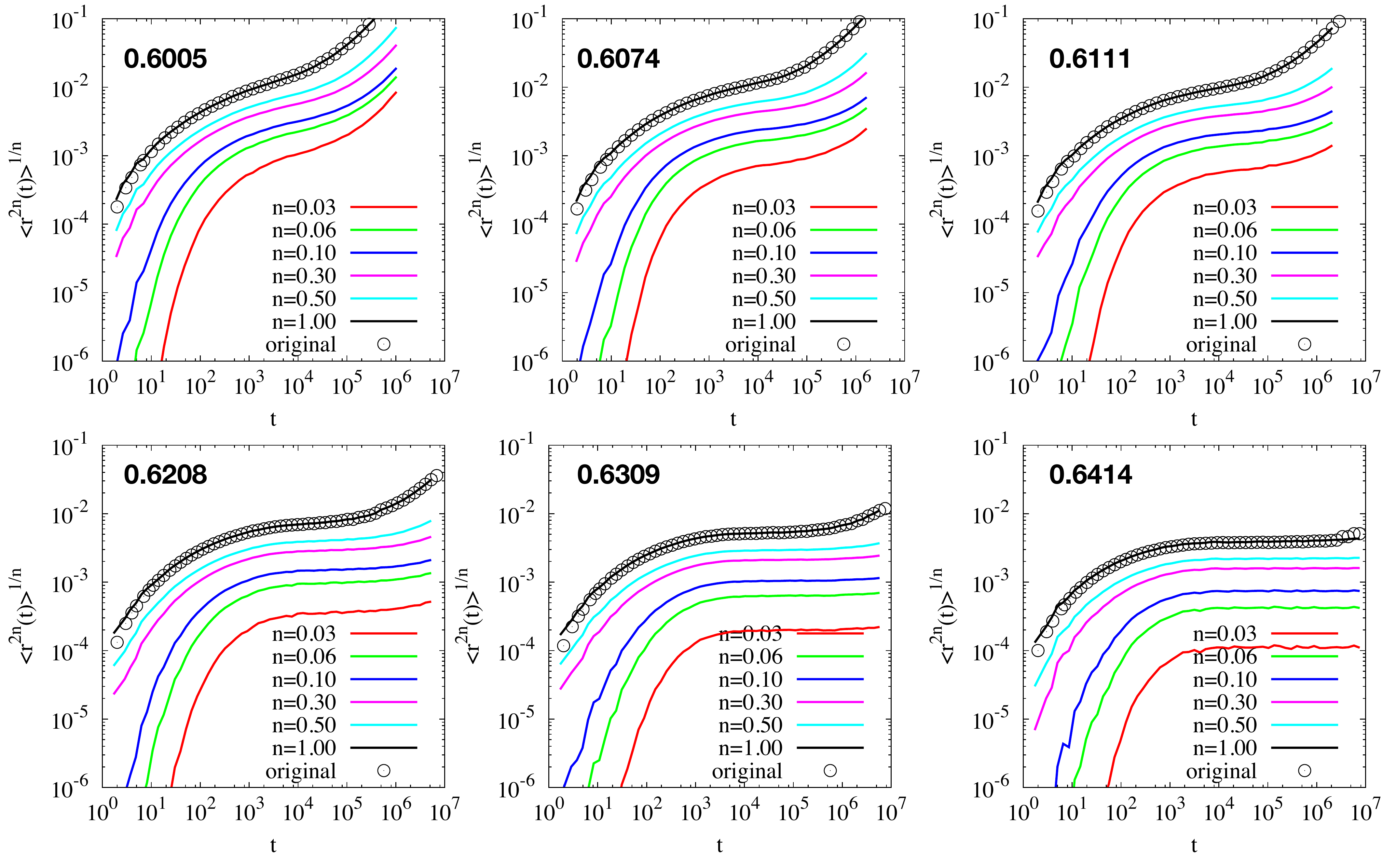}
\caption{Generalized mean square displacement for different $n$ and different $\varphi$ in $d=3$.}
\label{figS01}
\end{figure}

\begin{figure}
\includegraphics[width=\linewidth]{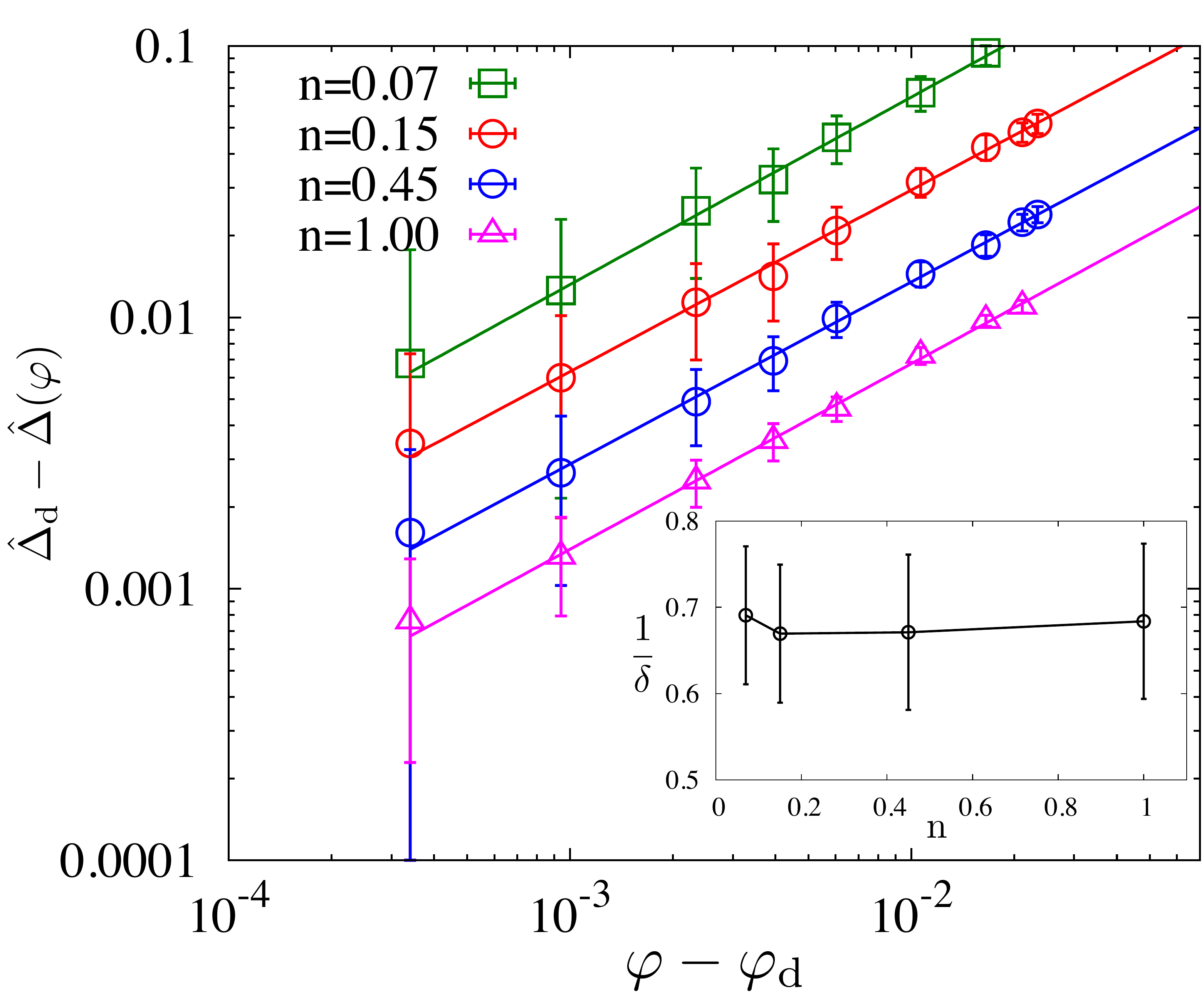}
\caption{Critical scaling of the typical cage size in $d=4$ calculated from the generalized MSD with packing fraction for different values of $n$. Inset: the critical exponent $\delta$ is robust against the choice of $n$. All data are for $N=1000$.}
\label{fit_gamma_4d}
\end{figure}
In order to further suppress the contribution of activated processes, we can generalize the definition of the mean square displacement as in Ref.~\c{patrick_pnas2014} 
\begin{equation}
\Delta r^2(t)=\lim_{n \to 0} \Bigg \{ \int r^{2n} G_{\rm s}(r,t) S_{d-1}(r) dr \Bigg \}^{1/n},
\label{msd}
\end{equation}
\begin{figure}
\includegraphics[width=\linewidth]{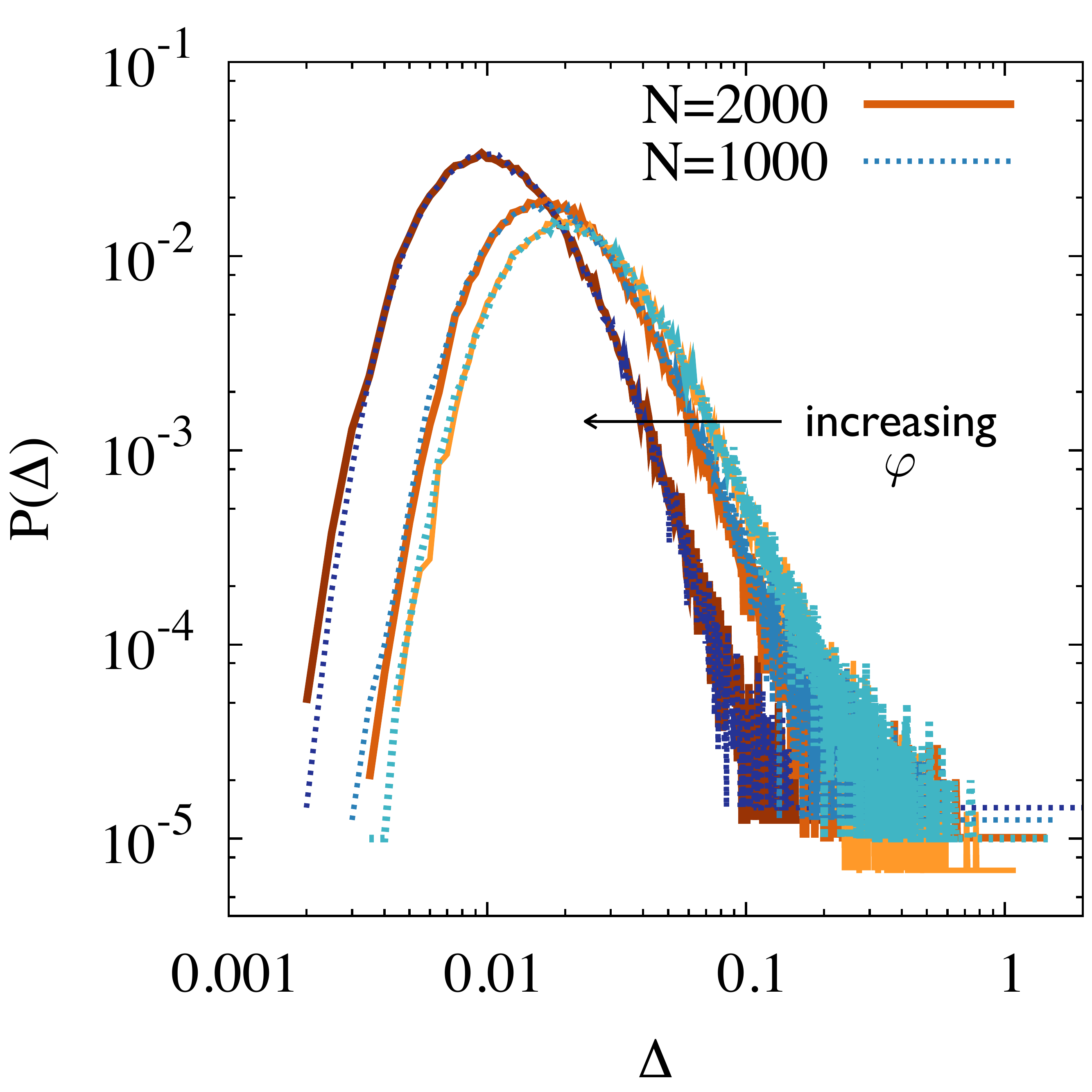}
\caption{Cage size distribution for $d=4$ at $\varphi=0.411, 0.416,$ and $0.431$ for different system sizes. The typical cage size is found to be robust against changes to the system size.}
\label{figS03}
\end{figure}
where $S_{d-1}(r)=2 \pi^{d/2}r^{d-1}/\Gamma(d/2)$ is the surface area of a $d$-dimensional hypersphere, and $G_\mathrm{s}({\bf r},t)$ is the self-part of the van Hove function. 
Note, however, that a direct use of this expression is numerically impractical, because it requires perfect sampling around the maximum of the van Hove function. In practice, we thus consider the convergence of the analysis upon reducing $n$ from the standard MSD definition with $n=1$. Figure~\ref{figS01}shows that reducing $n$ markedly enhances plateau formation, as expected. Thus, for smaller $n$, the cage size distribution can be measured over a longer time window, which is particularly helpful as $\varphi$ approaches $\varphi_{\rm d}$. For instance, with $n=0.07$ the upper bound of the time window can be set at $ 25\%$ (instead of $20\%$) of the peak position of $\alpha_2 (t)$. 
The critical exponent extracted from fitting results for different values of $n$ to Eq.~\ref{eq:cage} is robust. As an illustration, results for $d=4$ are reported in \f{fit_gamma_4d}. 

\section{Finite-size Effects}
\label{sec:finitesize}
To ensure the robustness of the critical scaling with system size, we consider the cage size distributions at three different densities above $\varphi_{\rm d}$ for two system sizes in \f{figS03}. The typical cage size or the mode of the distributions at a given density are here indistinguishable. The critical scaling of the typical cage is thus insensitive to the system size, in the regime considered here.

\end{document}